\documentclass[prd,aps,preprintnumbers, showpacs, nofootinbib,superscriptaddress,notitlepage]{revtex4-2}
\usepackage{slashed}
\usepackage{array}

\usepackage{multirow}

\usepackage{subfigure}

\usepackage{color}

\usepackage{epsfig}
\usepackage{bbold}

\usepackage{caption}

\usepackage{amssymb,amsthm,amsmath}
\usepackage{textcomp}
\usepackage{color}      
\usepackage{slashed}    
\usepackage{verbatim}
\usepackage[normalem]{ulem} 
\usepackage{rotating}   
\usepackage{multirow}   
\begin{document}
	\title{ Constraints on $\Delta L=2$ Vector Bosons with Tree Couplings to SM Particles	} 
	
		\author { Zhong-Lv Huang}\email{huangzhonglv@sjtu.edu.cn}

	\affiliation{
		State Key Laboratory of Dark Matter Physics,
		Tsung-Dao Lee Institute \& School of Physics and 
		Astronomy, Shanghai Jiao Tong University, Shanghai 200240, China} 
	\affiliation{Key Laboratory for Particle Astrophysics and Cosmology (MOE)
		\& Shanghai Key Laboratory for Particle Physics and Cosmology,
	Tsung-Dao Lee Institute  \&	School of Physics and Astronomy, Shanghai Jiao Tong University, Shanghai 200240, China}
	
		\author {Xiao-Gang He}\email{hexg@sjtu.edu.cn}
			
		\affiliation{
		State Key Laboratory of Dark Matter Physics,
		Tsung-Dao Lee Institute \& School of Physics and 
		Astronomy, Shanghai Jiao Tong University, Shanghai 200240, China} 
	\affiliation{Key Laboratory for Particle Astrophysics and Cosmology (MOE)
		\& Shanghai Key Laboratory for Particle Physics and Cosmology,
		Tsung-Dao Lee Institute  \&	School of Physics and Astronomy, Shanghai Jiao Tong University, Shanghai 200240, China}
	\date{\today}
	
\begin{abstract}
We investigate phenomenological implications of vector bosons $V$ transforming as (1, 2, -3/2) under the standard model (SM) product gauge group $SU(3)_C$, $SU(2)_L$ and $U(1)_Y$. These vector bosons can couple to two SM leptons at tree-level forming dimension-4 operators. These operators dictate $V$ to have two units of global lepton number, $\Delta L = 2$. The operators generated conserve the global lepton number but can violate generational lepton numbers. We study constraints on the couplings $Y$ of $V$ to SM particles using tree-level processes such as $l_\alpha^{-} \to l_\beta^{+} l_\rho^{-} l_\sigma^{-}$, muonium and antimuonium oscillation, neutrino trident scattering, inverse muon decay,  $e^- e^+ \to l^- l^+$, and also one-loop level processes such as the magnetic dipole moment of a charged lepton and $l_i \to l_j \gamma$. Strong constraints are obtained from $l_\alpha^{-} \to l_\beta^{+} l_\rho^{-} l_\sigma^{-}$ with $\left|Y_{e e} Y^{*}_{\mu e}\right| 
< 3.29 \times 10^{-11}\left(m_{V}/ \mathrm{GeV}\right)^{2},\left|Y_{e e} Y^{*}_{e \mu}\right| 
< 3.29 \times 10^{-11}\left(m_{V}/ \mathrm{GeV}\right)^{2}$ and from $l_i \to l_j \gamma$ with $ \left|Y_{\tau e}Y_{\mu\tau}^{*}\right|<3.46\times10^{-12}(m_{V}/ \mathrm{GeV})^2, \left|Y_{e \tau}Y_{\tau \mu}^{*}\right|<3.46\times10^{-12}\left(m_{V}/ \mathrm{GeV}\right)^2$, respectively. Interestingly, the imaginary part of the coupling constant in our model induces CP violation, which is constrained by experimental limits on the electric dipole moment. 
\end{abstract}

\maketitle

	\maketitle

\section{Introduction}
The Standard Model (SM) is very successful in accounting for the existing experimental data~\cite{ParticleDataGroup:2024cfk}. There are, however, hints suggesting the need for new interactions and particles from the neutrino and dark matter sectors, as well as anomalies associated with rare decays. There are several ways to introduce new physics, such as enlarging the gauge group for interactions and introducing new particles to the SM. Keeping the SM gauge group intact, the introduction of new particles in the SM can achieve the goal of modifying interactions. The new particles can be fermions~\cite{Minkowski:1977sc,Yanagida:1979as,Gell-Mann:1979vob,Glashow:1979nm,Mohapatra:1979ia,Foot:1988aq}, scalars~\cite{Magg:1980ut,Schechter:1980gr,Lazarides:1980nt,Davies:1990sc,Deshpande:1994vf,Bauer:2015knc,Crivellin:2017zlb}, or  vector particles~\cite{Singer:1980sw,Babu:1985gi,Montero:1992jk,Davidson:1993qk,Foot:1994ym,Fajfer:2015ycq,DiLuzio:2017vat,Cornella:2019hct,Cheung:2022zsb}, carrying different lepton and baryon numbers. There are many interesting phenomenological consequences associated with new particles and their properties can therefore be constrained using existing experimental data. In this work, we focus on new particles with two units of lepton number, $\Delta L =2$, which can couple directly to SM particles at tree-level.

There are four possible bi-lepton combinations which can lead to new particles with $\Delta L = 2$ that couple to SM leptons at tree-level. They are $\bar e_R e^c_R: (1, 1, 2)$, $\bar L_L L^c_L: (1, 1, 1) \oplus (1, 3, 1)$ and $\bar L_L \gamma_\mu e^c_R: (1, 2, 3/2)$. The new particles couple to these bi-lepton combinations corresponding to the first three combinations are scalars $k^{--}: (1, 1, - 2)$, $k^ -: (1, 1, -1)$ and $\chi = (\chi^{0}, \chi^-, \chi^{--})^T: (1, 3, -1)$. The last one is a vector doublet $V_\mu = (V^-_\mu, V^{--}_\mu)^T:  (1, 2, - 3/2)$. Here the numbers in the brackets are the corresponding $SU(3)_C$, $SU(2)_L$ and $U(1)_Y$ quantum numbers. The scalar particles with $\Delta L = 2$ interaction have been studied extensively in the construction of neutrino mass models, such as the Zee~\cite{Zee:1980ai}, Babu-Zee two loop neutrino mass~\cite{Zee:1985id,Babu:1988ki} and tree-level triplet neutrino mass models\cite{Magg:1980ut,Schechter:1980gr,Lazarides:1980nt}, as well as their other phenomenological consequences. The vector boson $V_\mu$ interactions with the SM particles have also been studied~\cite{delAguila:2010mx,Biggio:2016wyy}, but much less extensively compared with its scalar partners. We will make efforts to study phenomenological implications related to $V_\mu$ from several aspects, including the contributions of $V$ to leptons using data from tree-level processes such as $l_\alpha^{-} \to l_\beta^{+} l_\rho^{-} l_\sigma^{-}$, muonium and antimuonium oscillation, trident neutrino scattering $\nu_\mu N \to N + \mu \bar \mu \nu_i$, inverse muon decay, $e^- e^+ \to l^-l^+$, and loop-level processes such as $l_i \to l_j \gamma$, the magnetic dipole moment $g-2$ of charged leptons to obtain constraints for the relevant couplings. In particular, we will also examine the contribution to the electric dipole moment (EDM) of charged leptons from the doubly charged vector boson in our model, which also provides constraints on the relevant couplings.

The rest of the paper is organized as follows. We provide a detailed description of the model and the relevant interaction couplings in Section 2. In Section 3, we describe the impact of the vector particle on various observables used in our analysis, providing both the results and the allowed parameter space. We conclude in Section 4.

\section{Interactions of the new vectors with SM particles}
As mentioned before, the vector bosons $V$ transform as (1, 2, -3/2). Being a $SU(2)_L$ doublet, it has two component fields
\begin{eqnarray}
	V_\mu(1,2,-3/2)=\begin{pmatrix}V_\mu^-\\V_\mu^{--}\end{pmatrix}.
\end{eqnarray}
Their couplings to the SM bi-leptons with dimension-4 operators are as follows
\begin{eqnarray}
	\begin{aligned}
		\mathcal{L}=&Y_{\alpha\beta}\overline{L_{\alpha L}}\gamma^{\mu}l_{\beta R}^{c}V_{\mu}+\mathrm{h.c.}\\=&Y_{\alpha\beta}\left(\overline{\nu_{\alpha L}}\gamma^{\mu}l_{\beta R}^{c}V_{\mu}^{-}+\overline{l_{\alpha L}}\gamma^{\mu}l_{\beta R}^{c}V_{\mu}^{--}\right)+Y_{\alpha\beta}^{*}\left(\overline{l_{\beta R}^{c}}\gamma^{\mu}\nu_{\alpha L}V_{\mu}^{+}+\overline{l_{\beta R}^{c}}\gamma^{\mu}l_{\alpha L}V_{\mu}^{++}\right).
	\end{aligned} \label{f-v}
\end{eqnarray}
Here we have included generation indices $\alpha$ and $\beta$ in the Lagrangian. One cannot have dimension-4 operators where $V_\mu$ couples to SM quark fields and the Higgs field. 

The kinetic energy and mass terms of $V_\mu$, as well as its interactions with SM gauge bosons can be parameterized as~\cite{Baker:2019sli,Hiller:2021pul} 
\begin{eqnarray}
	\begin{aligned}
		\mathcal{L}_{V}=&-\frac{1}{2}V_{\mu\nu}^{\dagger}V^{\mu\nu}+M_{V}^{2}V_{\mu}^{\dagger}V^{\mu}
		-ig(1-\kappa_w)V_\mu^\dagger\frac{\tau^k}{2}V_\nu W^{k,\mu\nu}-ig^\prime(1-\kappa_y)V_\mu^\dagger YV_\nu B^{\mu\nu},
	\end{aligned}
\end{eqnarray}
where $M_V$ is the mass of $V_\mu$ and $V_{\mu\nu}=\sum_{i=1,2}D_{\mu}V_{\nu}^{i}-D_{\nu}V_{\mu}^{i}$ with $D_\mu=\partial_\mu-i\frac{g^{\prime}}2YB_\mu-i\frac g2\sigma_jW_\mu^j$. From the above Lagrangian,  the two components $V^+$ and $V^{++}$ masses $m_{V^{+}}=M_V$ and $m_{V^{++}}=M_V$ are degenerate. The values of $\kappa_w$ and $\kappa_y$ depend on the ultraviolet completion of the model. $B$ and $W$ are the $SU(2)_L$ and $U(1)_Y$ gauge fields, respectively. Using the relations between the $W$, $B$ fields, and the photon $A$ and $Z$ fields in the SM,
\begin{eqnarray}
	B^{\mu\nu}=\cos\theta_W F^{\mu\nu}-\sin\theta_WZ^{\mu\nu} , W^{3,\mu\nu}=\sin\theta_W F^{\mu\nu}+\cos\theta_WZ^{\mu\nu}, 
\end{eqnarray}
where $\theta_W$ is the weak mixing Weinberg angle. Then one can derive the coupling between $V$ and photon as
\begin{eqnarray}
	\begin{aligned}
		&\mathcal{L}_{V\gamma}=ie(1+\frac{1}{2}\kappa_w-\frac{3}{2}\kappa_y)V_\mu^{- \dagger}V_\nu^- F^{\mu\nu}+ ie(2-\frac{1}{2}\kappa_w-\frac{3}{2}\kappa_y)V_\mu^{-- \dagger}V_\nu^{--} F^{\mu\nu},
	\end{aligned}
\end{eqnarray}
where $F^{\mu\nu}=D_\mu A_\nu-D_\nu A_\mu $.

If we set $\kappa_w=\kappa_y=0$, the interaction Lagrangian becomes
\begin{eqnarray}
	\mathcal{L}_{V\gamma}=-\frac12V_{\mu\nu}^\dagger V^{\mu\nu}+ieQ_VV_\mu^\dagger V_\nu F^{\mu\nu},
\end{eqnarray}
where $D_{\mu}=\partial_{\mu}+ieQ_{V}A_{\mu}$.
Expanding $V^\dagger V A$ term, we obtain
\begin{eqnarray}
	\begin{aligned}\mathcal{L}_{V^{\dagger}VA}&=-\frac{1}{2}[ieQ_{V}(\partial_{\mu}V_{\nu}^{\dagger}-\partial_{\nu}V_{\mu}^{\dagger})(A^{\mu}V^{\nu}-A^{\nu}V^{\mu})-ieQ_{V}(A_{\mu}V_{\nu}^{\dagger}-A_{\nu}V_{\mu}^{\dagger})(\partial^{\mu}V^{\nu}-\partial^{\nu}V^{\mu})]\\
		&+ieQ_{V}V_{\mu}^{\dagger}V_{\nu}(\partial^{\mu}A^{\nu}-\partial^{\nu}A^{\mu})\\&=-ieQ_{V}(\partial_{\mu}V_{\nu}^{\dagger}-\partial_{\nu}V_{\mu}^{\dagger})A^{\mu}V^{\nu}+ieQ_{V}A_{\mu}V_{\nu}^{\dagger}(\partial^{\mu}V^{\nu}-\partial^{\nu}V^{\mu})+ieQ_{V}V_{\mu}^{\dagger}V_{\nu}(\partial^{\mu}A^{\nu}-\partial^{\mu}A^{\mu}).\end{aligned}
\end{eqnarray}
If  $\kappa_w=\kappa_y=1$, the new vectors will have no interaction with the SM gauge bosons. If $\kappa_w=\kappa_y=0$, the interaction of photon with $V$ is similar to its interaction with $W$ in the SM which may come from some full gauge theory~\cite{delAguila:2010mx,Biggio:2016wyy}.  If $\kappa_w$ and $\kappa_y$ take values different from the two cases, it will lead to divergences when carrying out loop calculations. Without a full theory, to avoid divergences in the subsequent calculations, we will adopt the condition $\kappa_w=\kappa_y=0$~\cite{Biggio:2016wyy}. 

In the following analysis, we will use experimental data to constrain the couplings described in the above.

\section{Phenomenological Constraints on the interactions}

We will use experimental data to study the constraints from tree-level processes, $l_\alpha^{-} \to l_\beta^{+} l_\rho^{-} l_\sigma^{-}$, $\nu_\mu N \to N + \mu \bar \mu \nu_i$, the muonium and antimuonium oscillation, inverse muon decay and $e^- e^+ \to l^- l^+$ processes, and then from loop-induced processes $l_i \to l_j \gamma$, $g-2$ and EDM of charged leptons.

\subsection{$l_\alpha^{-} \rightarrow l_\beta^{+} l_\rho^{-} l_\sigma^{-}$ interaction}

Exchanging $V^{--}_\mu$ at tree-level will generate $l_\alpha^{-} \rightarrow l_\beta^{+} l_\rho^{-} l_\sigma^{-}$ process,
\begin{eqnarray}
	\begin{aligned}
		\mathcal{L}=\frac{Y_{\rho\sigma}Y_{\alpha\beta}^*}{m_V^2}
		(\overline{l_{\rho}}\gamma_{\mu}P_L l_{\alpha})
		(\overline{l_{\sigma}}\gamma^{\mu}P_R l_{\beta}).
	\end{aligned}	\label{4l}
\end{eqnarray}
In the above, the flavor indices $\alpha,\beta,\rho,\sigma$ are summed over. For a given process $l_\alpha^{-} \rightarrow l_\beta^{+} l_\rho^{-} l_\sigma^{-}$, several operators contribute, as shown in the following equation,
\begin{eqnarray}
	\begin{aligned}
		M=\frac{Y_{\rho\sigma}Y_{\alpha\beta}^*}{m_V^2}(\overline{u(p_\rho)}\gamma_\mu P_Lu(p_\alpha))(\overline{u(p_\sigma)}\gamma^\mu P_Rv(p_\beta))-\frac{Y_{\rho\sigma}Y_{\beta\alpha}^*}{m_V^2}(\overline{u(p_\rho)}\gamma_\mu P_Lv(p_\beta))(\overline{u(p_\sigma)}\gamma^\mu P_Ru(p_\alpha))\\
		-\frac{Y_{\sigma\rho}Y_{\alpha\beta}^*}{m_V^2}(\overline{u(p_\sigma)}\gamma_\mu P_Lu(p_\alpha))(\overline{u(p_\rho)}\gamma^\mu P_Rv(p_\beta))+\frac{Y_{\sigma\rho}Y_{\beta\alpha}^*}{m_V^2}(\overline{u(p_\sigma)}\gamma_\mu P_Lv(p_\beta))(\overline{u(p_\rho)}\gamma^\mu P_Ru(p_\alpha))
	\end{aligned}
\end{eqnarray}
Then the initial particle spin averaged amplitude square will be

\begin{equation}
	\label{aml}
	\begin{aligned}
		\frac{1}{2}\sum_{s}|M|^2=&8(p_\alpha\cdot p_\sigma)(p_\beta\cdot p_\rho)\left( 
		\frac{|Y_{\rho\sigma}Y_{\alpha\beta}^*|^2}{m_V^4}
		+\frac{|Y_{\sigma\rho}Y_{\beta\alpha}^*|^2}{m_V^4}\right)\\
		&\quad+8(p_\alpha\cdot p_\rho)(p_\beta\cdot p_\sigma)\left( 
		\frac{|Y_{\rho\sigma}Y_{\beta\alpha}^*|^2}{m_V^4}
		+\frac{|Y_{\sigma\rho}Y_{\alpha\beta}^*|^2}{m_V^4}\right)\\
		&\quad+4m_{\alpha}m_{\beta}(p_{\rho}\cdot p_{\sigma})
		\frac{(Y_{\alpha\beta}^*Y_{\beta\alpha}+Y_{\alpha\beta}Y_{\beta\alpha}^*)(|Y_{\rho\sigma}|^2+|Y_{\sigma\rho}|^2)}{m_V^4}\\
		&\quad-4m_{\rho}m_{\sigma}(p_{\alpha}\cdot p_{\beta})
		\frac{(Y_{\rho\sigma}^*Y_{\sigma\rho}+Y_{\rho\sigma}Y_{\sigma\rho}^*)(|Y_{\beta\alpha}|^2+|Y_{\alpha\beta}|^2)}{m_V^4}\\
		&\quad-8m_{\alpha}m_{\beta}m_{\rho}m_{\sigma}\frac{(Y_{\rho\sigma}^*Y_{\sigma\rho}+Y_{\rho\sigma}Y_{\sigma\rho}^*)(Y_{\alpha\beta}^*Y_{\beta\alpha}+Y_{\alpha\beta}Y_{\beta\alpha}^*)}{m_V^4}.
	\end{aligned}
\end{equation}
Using the approximation condition $m_\alpha\gg m_\beta,m_\rho, m_\sigma$ and assuming that all coupling coefficients are in the same order, we can ignore the last three lines of Eq. (\ref{aml}) and the final result of the process will be
\begin{eqnarray}
	\Gamma\approx\frac{1}{1+\delta_{\rho\sigma}}\frac{1}{8}\frac{m_{\alpha}^{5}}{192\pi^{3}}\left(\frac{|Y_{\rho\sigma}Y_{\alpha\beta}^{*}|^{2}}{m_{V}^{4}}+\frac{|Y_{\sigma\rho}Y_{\beta\alpha}^{*}|^{2}}{m_{V}^{4}}+\frac{|Y_{\rho\sigma}Y_{\beta\alpha}^{*}|^{2}}{m_{V}^{4}}+\frac{|Y_{\sigma\rho}Y_{\alpha\beta}^{*}|^{2}}{m_{V}^{4}}\right).
\end{eqnarray}
Here the Kronecker delta $\delta_{\rho\sigma}$ takes the value 1 for possible two identical final states and the value 0 for different final states.
The result above will lead to tree-level decays of the types:  $\mu^- \to e^+e^-e^-$, $\tau^- \to e^+e^-e^-$, $\tau^- \to e^+e^-\mu^-$, $\tau^- \to e^+\mu^-\mu^-$, $\tau^- \to \mu^+e^-e^-$, $\tau^- \to \mu^+\mu^-e^-$, $\tau^- \to \mu^+\mu^-\mu^-$. Therefore, the corresponding coefficients are constrained by the experimental data of these rare lepton number violation processes~\cite{SINDRUM:1987nra,Hayasaka:2010np}.

\begin{table}[htbp] 
	\begin{center}
		\begin{tabular}{lcc}
			\hline\hline
			Process & Branching ratio bound & Constraints \\
			\hline 
			$\mu^- \to e^+e^-e^-$ & $1.0\times 10^{-12}$~\cite{SINDRUM:1987nra} & $\left|Y_{e e} Y^{*}_{\mu e}\right| 
			< 3.29 \times 10^{-11}\left(\frac{m_{V}}{ \mathrm{GeV}}\right)^{2},\left|Y_{e e} Y^{*}_{e \mu}\right| 
			< 3.29 \times 10^{-11}\left(\frac{m_{V}}{ \mathrm{GeV}}\right)^{2}$\\
			$\tau^- \to e^+e^-e^-$ & $2.7\times 10^{-8}$~\cite{Hayasaka:2010np} & $\left|Y_{e e} Y^{*}_{\tau e}\right| 
			< 1.28 \times 10^{-8}\left(\frac{m_{V}}{ \mathrm{GeV}}\right)^{2},\left|Y_{e e} Y^{*}_{e \tau}\right| 
			< 1.28 \times 10^{-8}\left(\frac{m_{V}}{ \mathrm{GeV}}\right)^{2}$\\
			$\tau^- \to e^+e^-\mu^-$ & $1.8\times 10^{-8}$~\cite{Hayasaka:2010np} & $\begin{cases}
				\left|Y_{e \mu} Y^{*}_{\tau e}\right| 
				< 1.05 \times 10^{-8}\left(\frac{m_{V}}{ \mathrm{GeV}}\right)^{2},\left|Y_{\mu e} Y^{*}_{e \tau}\right| 
				< 1.05 \times 10^{-8}\left(\frac{m_{V}}{ \mathrm{GeV}}\right)^{2}\\
				\left|Y_{e \mu} Y^{*}_{e \tau}\right| 
				< 1.05 \times 10^{-8}\left(\frac{m_{V}}{ \mathrm{GeV}}\right)^{2},\left|Y_{\mu e} Y^{*}_{\tau e}\right| 
				< 1.05 \times 10^{-8}\left(\frac{m_{V}}{ \mathrm{GeV}}\right)^{2}
			\end{cases}$\\
			$\tau^- \to e^+\mu^-\mu^-$ & $1.7\times10^{-8}$~\cite{Hayasaka:2010np} & $\left|Y_{\mu \mu} Y^{*}_{\tau e}\right| 
			< 1.02 \times 10^{-8}\left(\frac{m_{V}}{ \mathrm{GeV}}\right)^{2},\left|Y_{\mu \mu} Y^{*}_{e \tau}\right| 
			< 1.02 \times 10^{-8}\left(\frac{m_{V}}{ \mathrm{GeV}}\right)^{2} $\\
			$\tau^- \to \mu^+e^-e^-$ & $1.5\times 10^{-8}$~\cite{Hayasaka:2010np} &$	\left|Y_{ee} Y^{*}_{\tau \mu}\right| 
			< 9.57 \times 10^{-9}\left(\frac{m_{V}}{ \mathrm{GeV}}\right)^{2},\left|Y_{e e} Y^{*}_{\mu \tau}\right| 
			< 9.57 \times 10^{-9}\left(\frac{m_{V}}{ \mathrm{GeV}}\right)^{2} $ \\
			$\tau^- \to \mu^+\mu^-e^-$ & $2.7\times 10^{-8}$~\cite{Hayasaka:2010np} & $\begin{cases}
				\left|Y_{\mu e} Y^{*}_{\tau \mu}\right| 
				< 1.28 \times 10^{-8}\left(\frac{m_{V}}{ \mathrm{GeV}}\right)^{2},\left|Y_{e \mu} Y^{*}_{\mu \tau}\right| 
				< 1.28 \times 10^{-8}\left(\frac{m_{V}}{ \mathrm{GeV}}\right)^{2}\\
				\left|Y_{\mu e} Y^{*}_{\mu \tau}\right| 
				< 1.28 \times 10^{-8}\left(\frac{m_{V}}{ \mathrm{GeV}}\right)^{2},\left|Y_{e \mu} Y^{*}_{\tau \mu}\right| 
				< 1.28 \times 10^{-8}\left(\frac{m_{V}}{ \mathrm{GeV}}\right)^{2}
			\end{cases}$ \\
			$\tau^- \to \mu^+\mu^-\mu^-$ & $2.1\times 10^{-8}$~\cite{Hayasaka:2010np} & $	\left|Y_{\mu \mu} Y^{*}_{\tau \mu}\right| 
			< 1.13 \times 10^{-8}\left(\frac{m_{V}}{ \mathrm{GeV}}\right)^{2},\left|Y_{\mu \mu } Y^{*}_{\mu \tau}\right| 
			< 1.13 \times 10^{-8}\left(\frac{m_{V}}{ \mathrm{GeV}}\right)^{2} $\\
			\hline\hline
		\end{tabular}
		\caption{Constraints for the coupling and mass from different $l_\alpha^{-} \rightarrow l_\beta^{+} l_\rho^{-} l_\sigma^{-}$ processes. The second column shows the existing experimental limits at 90$\%$ CL. All the constraints in the table for the couplings are obtained under the assumption that only the corresponding parameters are non-zero and are shown in the third column.}
		\label{ltolll}
	\end{center}
\end{table}
In Table \ref{ltolll}, we show the constraints from $l_\alpha^{-} \rightarrow l_\beta^{+} l_\rho^{-} l_\sigma^{-}$. Experimentally there is no evidence to show that such decays exist. Therefore, one can use experimental data to constrain the couplings.  We show the 90$\%$ CL constraints in the second column of the table for the upper bound of certain combinations of coupling coefficients. The bounds for these flavor-violating processes are shown in the third column of the table. All restrictions here apply to the case where only the corresponding parameter contribution is considered to be non-zero. This also applies to the bounds presented in the rest of the paper. One can see from Table \ref{ltolll} that in this class of process, $\mu^- \to e^+e^-e^-$ gives the strongest constraints with $\left|Y_{e e} Y^{*}_{\mu e}\right| < 3.29 \times 10^{-11}\left(m_{V}/\mathrm{GeV}\right)^{2},\left|Y_{e e} Y^{*}_{e \mu}\right| < 3.29 \times 10^{-11}\left(m_{V}/ \mathrm{GeV}\right)^{2}$. However, it should be noted that the constraints obtained from different three-body charged lepton decays correspond to different combinations of $YY^{*}$, which means all the limits in the table are physically meaningful.

\subsection{Muonium and antimuonium oscillation}
Exchanging $V^{--}_\mu$ as described in Eq. (\ref{4l}) generates an interaction that can also induce muonium and antimuonium oscillation,
\begin{eqnarray}
	\label{muoniumL}
	\mathcal{L}_{M-\bar M}={Y_{\mu \mu} Y^*_{e e}\over m^2_V} \overline {\mu_{L}}\gamma^{\mu} e_L
	\overline {\mu_R}\gamma_{\mu}e_R,
	\label{m-m}
\end{eqnarray}
where muonium is a non-relativistic QED bound state of an antimuon and an electron. muonium and antimuonium oscillation is a $\Delta L_{\mu}=2$ process. If this phenomenon were observed in the current series of experiments, it would provide compelling evidence for the existence of new physics~\cite{Willmann:1998gd}.

Since our new physics Lagrangian includes lepton-flavor-violating interactions characterized by $\Delta L_\mu=2$, we can analyze combined evolution of muonium and antimuonium.
\begin{eqnarray}
	|\psi(t)\rangle=\left(\begin{array}{c}a(t)\\b(t)\end{array}\right)=a(t)|M_{\mu}\rangle+b(t)|\overline{M}_{\mu}\rangle.
\end{eqnarray}
The time evolution can be described by a Schrodinger equation
\begin{eqnarray}
	i\frac{d}{dt}\binom{\left|M_{\mu}\left(t\right)\right\rangle}{\left|\overline{M}_{\mu}\left(t\right)\right\rangle}=\left(m-i\frac{\Gamma}{2}\right)\binom{\left|M_{\mu}\left(t\right)\right\rangle}{\left|\overline{M}_{\mu}\left(t\right)\right\rangle},
\end{eqnarray}
where $m$ and $\Gamma$ are both $2\times 2$ hermitian matrices. The off-diagonal elements of the matrix can be written by
\begin{eqnarray}
	\label{muonium}
	\left(m-\frac{i}{2}\Gamma\right)_{12}=\frac{1}{2M_{M}}\langle\overline{M}_{\mu}|\mathcal{H}_{{\mathrm{eff}}}|M_{\mu}\rangle+\frac{1}{2M_{M}}\sum_{n}\frac{\langle\overline{M}_{\mu}|\mathcal{H}_{{\mathrm{eff}}}|n\rangle\langle n|\mathcal{H}_{{\mathrm{eff}}}|M_{\mu}\rangle}{M_{M}-E_{n}+i\epsilon},
\end{eqnarray}
where $m_{12}$ and $\Gamma_{12}$ contribute to the mass and lifetime differences between the two physical states of muonium, $\mathcal{H}_{{\mathrm{eff}}}=-\mathcal{L}_{M-\bar M}$ and in the second term, $n$ represents the possible intermediate states.

In our model, the new vector boson we introduced is expected to have a mass characteristic of new physics, which is significantly larger than the mass of the muonium. As a result, the contribution to $\Delta \Gamma$ is negligible, allowing us to focus solely on the mass difference~\cite{Chen:2022gmk}. Furthermore, CPT invariance ensures that the masses and decay widths of muonium and antimuonium are identical, implying $m_{11}=m_{22}$ and $\Gamma_{11}=\Gamma_{22}$.
The mass eigenstates after diagonalization can be defined as$|M_{\mu_{1,2}}\rangle=\frac1{\sqrt{2}}\left[|M_\mu\rangle\mp|\overline{M}_\mu\rangle\right]$, and we can define $\Delta m\equiv M_1-M_2, \Delta\Gamma\equiv\Gamma_2-\Gamma_1$ to be the mass and width differences of the mass eigenstates, respectively. Then we can derive the oscillation probability~\cite{Conlin:2020veq,Petrov:2022wau}
\begin{eqnarray}
	P(M_\mu\to\bar{M_\mu})=\frac12\left(x^2+y^2\right)\approx\frac{x^2}{2},
\end{eqnarray}
where the two dimensionless quantities $x=\Delta m/\Gamma, y=\Delta\Gamma / (2\Gamma)$ and the average lifetime $\Gamma=(\Gamma_1+\Gamma_2)/2$. From Eq. (\ref{muonium}), $x$ can be expressed as
\begin{eqnarray}
	x\approx\frac{1}{M_{M}\Gamma}\operatorname{Re}(\langle\overline{M}_\mu|\mathcal{H}_{\mathrm{eff}}|M_\mu\rangle).
\end{eqnarray}

Using the interaction we introduced from Eq. (\ref{muoniumL}), consider the spin-0 para-muonium contribution $x_P$ and spin-1 ortho-muonium contribution $x_V$ in the experiment, one can gain
\begin{eqnarray}
	x_P=\frac{24(m_{red}\alpha)^3}{\pi\Gamma}\frac{Y_{\mu \mu}Y_{e e}^{*}}{m_{V}^{2}},	x_V=\frac{72(m_{red}\alpha)^3}{\pi\Gamma}\frac{Y_{\mu \mu}Y_{e e}^{*}}{m_{V}^{2}},
\end{eqnarray}
where $ \varphi(0)$ is the muonium wave function at the origin with
\begin{eqnarray}
	|\varphi(0)|^2=\frac{(m_{\mathrm{red}}\alpha)^3}{\pi}=\frac{1}{\pi}(m_{\mathrm{red}}\alpha)^3,
\end{eqnarray}
and $m_{\mathrm{red}}=m_em_\mu/(m_\mu+m_e)\simeq m_e$ is the reduced mass of muonium, $\alpha$ is the fine structure constant.
We should also average the oscillation probability over the number of polarization degrees of freedom~\cite{Willmann:1998gd} to get the experimental oscillation probability
\begin{eqnarray}
	\begin{aligned}
		P(M_\mu\to\overline{M}_\mu)^{\mathrm{exp}}&=\sum_{i=P,V}\frac{1}{2s_i+1}P(M_\mu^i\to\overline{M}_\mu^i)\\
		&=\frac{1}{2}x_{P}^{2}+\frac{1}{6}x_{V}^{2}=\frac{24 (m_{red}\alpha)^6}{\pi^2\Gamma^2}\frac{|Y_{\mu \mu}Y_{e e}|^2}{m_{V}^{4}}.
	\end{aligned}
\end{eqnarray}

We shall use data from the most recent experiment to constrain the oscillation parameters. Considering the effect of the magnetic field in the experiment~\cite{Willmann:1998gd}, the upper bound on muonium and antimuonium oscillation was obtained as
\begin{eqnarray}
	P(M_\mu\to\overline{M}_\mu)^{\exp}\leq8.3\times10^{-11}/S_B(B_0),
\end{eqnarray}
at the 90$\%$ CL, where $S_B(B_0)=0.95$~\cite{Willmann:1998gd}. So, the final constraint on the coupling and mass of the new vector doublet is
\begin{eqnarray}
	|Y_{\mu \mu}Y_{e e}^{*}|<3.52\times10^{-8} \left(\frac{m_{V}}{ \mathrm{GeV}}\right)^{2}.
\end{eqnarray}
The constraint given above is slightly weaker than those given in the previous section. However, this corresponds to a different combination of coupling coefficients compared to the constraints in Table \ref{ltolll}, so it may still be useful. 

\subsection{Neutrino Trident Scattering}

The new vector will also contribute to the neutrino trident scattering process,  $\nu_\mu N \to N + \mu \bar \mu \nu_i$. Since the neutrino trident scattering process is highly sensitive to the new physics~\cite{Altmannshofer:2014pba,Altmannshofer:2016brv,  Cen:2021iwv, Cheng:2021okr}, it can play an important role in constraining new physics beyond SM.
From Eq. (\ref{f-v}), the new contribution is given by
\begin{eqnarray}
	\mathcal{L}=\frac{Y_{\rho\sigma}Y_{\alpha\beta}^{*}}{m_{V}^{2}}(\overline{\nu_{\rho}}\gamma_{\mu}P_{L}\nu_{\alpha})(\overline{l_{\sigma}}\gamma^{\mu}P_{R}l_{\beta}).
\end{eqnarray}
This effective interaction will contribute to the neutrino trident scattering process and lead to a modification of the measured $\sigma/\sigma_{\mathrm{SM}}$, where the ratio is the cross section $\sigma$ from experimental measurements to the SM predicted cross section $\sigma_{SM}$. If there are no corrections to the SM contribution, the ratio equals 1. Compared to the SM contribution, we should also sum over all the final state of the neutrino as the flavors of the final neutrinos are not identified in the experiment. Therefore, the theoretical expression with new physics contribution for the ratio $\sigma_{new}/\sigma_{SM}$ can be written by
\begin{eqnarray}
	\frac{\sigma_{new}}{\sigma_{\mathrm{SM}}}=\frac{\left(1+4 s_{w}^{2}-\frac{4m_w^2}{g^2} \frac{\left|Y_{\mu\mu}\right|^{2}}{m_{V}^{2}}\right)^{2}+\left(1+\frac{4m_w^2}{g^2} \frac{\left|Y_{\mu\mu}\right|^{2}}{m_{V}^{2}}\right)^2+2\left(\frac{4m_w^2}{g^2} \frac{\left|Y_{\mu\mu}\right|^{2}}{m_{V}^{2}}\right)^2\left(\frac{\left|Y_{e \mu}\right|^2+\left|Y_{\tau \mu}\right|^2}{\left|Y_{\mu\mu}\right|^{2}}\right)}{\left(1+4 s_{w}^{2}\right)^{2}+1}.
	\label{root}
\end{eqnarray}
From the equation, we can see that two terms,  ${\left|Y_{\mu\mu}\right|^{2}}/{m_{V}^{2}}$ and ${(\left|Y_{e \mu}\right|^2+\left|Y_{\tau \mu}\right|^2)}/{m_{V}^{2}}$, will have an influence on the cross section driving the ratio away from 1. And the experimental data of the ratio are $1.58\pm 0.57$~\cite{CHARM-II:1990dvf}, $0.82\pm 0.28$~\cite{CCFR:1991lpl} and $0.72^{+1.73}_{-0.72}$~\cite{NuTeV:1999wlw}, respectively, which lead to the average value $\sigma_{\mathrm{exp}}/\sigma_{\mathrm{SM}}=0.95\pm0.25$.

\begin{figure*}[htbp] 
	\centering
	\includegraphics[width=4in]{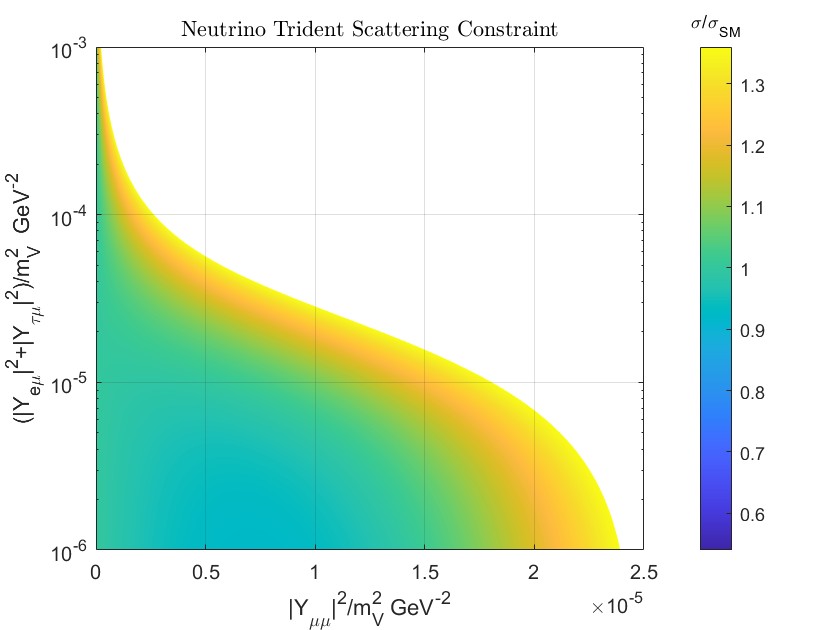}
	\caption{\label{trident}Neutrino trident scattering constraint on the coupling constants. The colored region is the allowed parameter space from the experiment at $90\%$ CL level.
	}
\end{figure*}
We also show the allowed range for $\sigma_{new}/\sigma_{\mathrm{SM}}$ from neutrino trident scattering in Figure \ref{trident}. The colored region represents the parameter space allowed by the experiment at $90\%$ CL. From Figure \ref{trident} we can see that the experimental values we predict will revert to the standard model results when our model interaction does not exist, and also provide a constraint
\begin{eqnarray}
	{\left|Y_{\mu\mu}\right|^{2}}<2.46\times10^{-5}\left(\frac{m_{V}}{ \mathrm{GeV}}\right)^{2}.
\end{eqnarray}

However, as ${\left|Y_{\mu\mu}\right|^{2}}/{m_{V}^{2}}$ approaching 0, the constraint on $\left|Y_{e \mu}\right|^2+\left|Y_{\tau \mu}\right|^2$ is weak as shown in Figure~\ref{trident}. This is because the contribution of $\left|Y_{e \mu}\right|^2+\left|Y_{\tau \mu}\right|^2$ is suppressed by the possibly extremely small $\left|Y_{\mu\mu}\right|^{2}$ from the third term in the numerator of Eq. (\ref{root}). 

\subsection{Inverse muon decay}
The study of neutrino-lepton interactions, not only neutrino trident scattering, but also inverse muon decay provides valuable insights into the constraints on our model.
The CHARM \cite{Amsterdam-CERN-Hamburg-Moscow-Rome:1980qbb} and NuTeV \cite{NuTeV:2001bgq} collaborations investigated inverse muon decay process, $\nu_\mu e^-\to\mu^-\nu_e$, and also the lepton number violating process $\bar{\nu}_\mu e^-\to\mu^-\bar{\nu}_e$, where the SM has no contribution to the second channel. The upper limit for the ratio of these processes is
\begin{eqnarray}
	\frac{\sigma\left(\bar{\nu}_\mu e^{-} \rightarrow \mu^{-} \bar{\nu}_e\right)}{\sigma\left(\nu_\mu e^{-} \rightarrow \mu^{-} \nu_e\right)}<\left\{\begin{array}{l}
		0.09 \text{~\cite{Amsterdam-CERN-Hamburg-Moscow-Rome:1980qbb}}\\
		0.017 \text{~\cite{NuTeV:2001bgq}}
	\end{array}\right.
\end{eqnarray}
It is evident that our new vector boson can contribute to the lepton number violating process $\bar{\nu}_\mu e^-\to\mu^-\bar{\nu}_e$,
\begin{eqnarray}
	\begin{aligned}
		M=\frac{Y_{\mu \mu}Y_{e e}^*}{m_V^2}(\overline{v(p_{\nu_\mu})}\gamma_\mu P_Lv(p_{\mu_e}))(\overline{u(p_\mu)}\gamma^\mu P_Ru(p_e)).
	\end{aligned}
\end{eqnarray}
So we can obtain the ratio of these processes
\begin{eqnarray}
	\frac{\sigma(\bar{\nu}_\mu e^-\to\mu^-\bar{\nu}_e)}{\sigma(\nu_\mu e^-\to\mu^-\nu_e)}=\frac{|Y_{\mu\mu}Y^*_{ee}|^2}{8 G_F^2 m_V^4},
\end{eqnarray}
where we neglect the new physics contribution to $\bar{\nu}_\mu e^-\to\mu^-\bar{\nu}_e$ as it is much smaller than the SM contribution. Thus, the constraints can be derived as
\begin{eqnarray}
	\left|Y_{\mu \mu} Y_{e e}^*\right|<\left\{\begin{array}{l}
		9.90 \times 10^{-6}\left(\frac{m_V}{\mathrm{GeV}}\right)^2 \text{~\cite{Amsterdam-CERN-Hamburg-Moscow-Rome:1980qbb}}\\
		4.30 \times 10^{-6}\left(\frac{m_V}{\mathrm{GeV}}\right)^2 \text{~\cite{NuTeV:2001bgq}}
	\end{array}\right.
\end{eqnarray}
which provide different limits on the different combination of coupling coefficients.

\subsection{$e^-e^+ \to l^-l^+$}
The new interactions introduced will affect $e^-e^+ \to l^-l^+$ scattering processes.
We can also obtain constraints on the new interactions from precise lepton scattering processes in colliders such as LEP II. 

The interaction in Eq. (\ref{4l}) will interfere with the SM contribution to $e^-e^+ \to l^-l^+$ process. Considering the constructive or destructive interference with the Standard Model and limiting to leading order, and assuming that the vector mass is much bigger than the center-of-mass energy at LEP II, the additional term of the amplitude beyond the SM of $e_\alpha^-e_\beta^+ \to e_\rho^-e_\sigma^+$ at leading order will be
{\footnotesize
	\begin{eqnarray}
		\label{eeee}
		\begin{aligned}
			\frac14\sum_s|M|^2_{\text{NP}}=& 4(p_{\alpha}\cdot p_{\rho})(p_{\beta}\cdot p_{\sigma})\left(\frac{|Y_{\beta\rho}Y_{\alpha\sigma}^*|^2}{m_{V}^4}+\frac{|Y_{\rho\beta}Y_{\sigma\alpha}^*|^2}{m_{V}^4}\right) +4(p_\alpha\cdot p_\beta)(p_\rho\cdot p_\sigma)\left(\frac{|Y_{\rho\beta}Y_{\alpha\sigma}^*|^2}{m_V^4}+\frac{|Y_{\beta\rho}Y_{\sigma\alpha}^*|^2}{m_V^4}\right) \\
			+&\frac{2e^2 (Y_{\alpha\sigma}^*Y_{\beta\rho}+Y_{\alpha\sigma}Y_{\beta\rho}^*+Y_{\sigma\alpha}Y_{\rho\beta}^*+Y_{\sigma\alpha}^*Y_{\rho\beta})\left(m_l^2(g_A^2+g_V^2)(p_{\alpha}\cdot p_{\beta})+2(p_{\alpha}\cdot p_{\rho})(p_{\beta}\cdot p_{\sigma})\right)}{\mathrm{m}_V^2q^2}\\
			+&\frac{(-Y_{\alpha\sigma}^*Y_{\beta\rho}- Y_{\alpha\sigma}Y_{\beta\rho}^* ) (m_l^2(g_A^2+g_V^2)(p_{\alpha}\cdot p_{\beta})+2(g_V^2-g_A^2)(p_{\alpha}\cdot p_{\rho})(p_{\beta}\cdot p_{\sigma})  )       }{2m_V^2(m_z^2-q^2)}\\
			+&\frac{(-Y_{\sigma\alpha}^*Y_{\rho\beta}- Y_{\sigma\alpha}Y_{\rho\beta}^* ) (m_l^2(g_A^2-g_V^2)(p_{\alpha}\cdot p_{\beta})+2(g_V^2-g_A^2)(p_{\alpha}\cdot p_{\rho})(p_{\beta}\cdot p_{\sigma})  )       }{2m_V^2(m_z^2-q^2)}\\
			-&\frac{2e^2 (Y_{\alpha\sigma}^*Y_{\rho\beta}+Y_{\sigma\alpha}Y_{\beta\rho}^*+Y_{\alpha\sigma}Y_{\rho\beta}^*+Y_{\sigma\alpha}^*Y_{\beta\rho})\left(2(p_{\alpha}\cdot p_{\beta})(p_{\rho}\cdot p_{\sigma})\right)}{\mathrm{m}_V^2q_t^2}\\
			+&\frac{(g_A^2-g_V^2)  (p_{\alpha}\cdot p_{\beta})(p_{\rho}\cdot p_{\sigma}) (-Y_{\alpha\sigma}^*Y_{\rho\beta}-Y_{\rho\beta}^*Y_{\alpha\sigma}   )  }{m_V^2(m_z^2-q_t^2)}\\
			+&\frac{(g_A^2-g_V^2)  (p_{\alpha}\cdot p_{\beta})(p_{\rho}\cdot p_{\sigma}) (-Y_{\beta\rho}^*Y_{\sigma\alpha}-Y_{\sigma\alpha}^*Y_{\beta\rho}    )  }{m_V^2(m_z^2-q_t^2)}
		\end{aligned}
	\end{eqnarray}}
where $q$ is the momentum transfer in the s channel, and $q_t$ is the momentum transfer in the t channel. $m_l$ is the mass of final state particle, which can be neglected here because the mass of the electron is much smaller than the center-of-mass energy of the experiment when the final-state particles are electrons. It is clear that the new physics, along with its interference with the Standard Model, will affect the amplitude.

When considering $e^-e^+ \to \mu^-\mu^+$ or $e^-e^+ \to \tau^- \tau^+$ process, one can obtain the formula by removing the last three lines of Eq. (\ref{eeee}) since there are no additional t-channel contribution. Note that $m_l$ should be taken as the muon or tauon mass.

\begin{table}[htbp] 
	\begin{center}
		\begin{tabular}{lcl}
			\hline\hline
			Process & $\sigma_{exp}/\sigma_{SM}\pm(stat)\pm(syst)\pm(theory)$  &\multicolumn{1}{c}{Constraints}\\
			\hline 
			$e^-e^+ \to e^-e^+$ & $1.0006\pm0.0086\pm0.0077\pm0.0200$ & $
			~~{Y_{ee}Y_{ee}^*}<6.10\times10^{-6}\left(\frac{m_{V}}{ \mathrm{GeV}}\right)^{2}$ \\
			$e^-e^+ \to \mu^-\mu^+$ & $0.9961\pm0.0244\pm0.0062\pm0.0040$ & $
			\begin{cases}
				Y_{e\mu}Y_{e\mu}^*<5.61\times10^{-7}\left(\frac{m_{V}}{ \mathrm{GeV}}\right)^{2},{Y_{\mu e}Y_{\mu e}^*}<1.97\times10^{-7}\left(\frac{m_{V}}{ \mathrm{GeV}}\right)^{2}\\
				{|Y_{\mu e}Y_{e\mu}^*|}<1.36\times10^{-6}\left(\frac{m_{V}}{ \mathrm{GeV}}\right)^{2},{|Y_{e \mu}Y_{\mu e}^*|}<1.36\times10^{-6}\left(\frac{m_{V}}{ \mathrm{GeV}}\right)^{2}
			\end{cases}$ \\
			$e^-e^+ \to \tau^-\tau^+$  & $0.9852\pm0.0341\pm0.0203\pm0.0040$ &$\begin{cases}
				{Y_{e\tau}Y_{e\tau}^*}<7.28\times10^{-7}\left(\frac{m_{V}}{ \mathrm{GeV}}\right)^{2},{Y_{\tau e}Y_{\tau e}^*}<2.68\times10^{-7}\left(\frac{m_{V}}{ \mathrm{GeV}}\right)^{2}\\
				{|Y_{\tau e}Y_{e \tau}^*|}<1.59\times10^{-6}\left(\frac{m_{V}}{ \mathrm{GeV}}\right)^{2},{|Y_{e\tau}Y_{\tau e}^*|}<1.59\times10^{-6}\left(\frac{m_{V}}{ \mathrm{GeV}}\right)^{2}
			\end{cases} $\\
			\hline\hline
		\end{tabular}
		\caption{$e^-e^+ \to l^- l^+$ limit from LEP II. The second column shows the existing experimental limits and the errors are the statistical, experimental systematic and theoretical uncertainties. The corresponding  bounds are shown in the third column of the table at $90\%$ CL level.}
		\label{llll}
	\end{center}
\end{table}
As discussed earlier, the coupling coefficients of the new physics are constrained by the LEP II data~\cite{DELPHI:2005wxt}. All the corresponding restrictions can be found in Table. \ref{llll}, and one can see these constraints are generally weak in comparison to the previous ones. However, they correspond to different parameter spaces of the coupling coefficients. Therefore, even the seemingly weaker constraints have physical significance.

\subsection{$l_i \to l_j \gamma$, magnetic dipole moment and electric dipole moment of charged leptons}
\begin{figure*}[htbp] 
	\centering
	\includegraphics[width=4in]{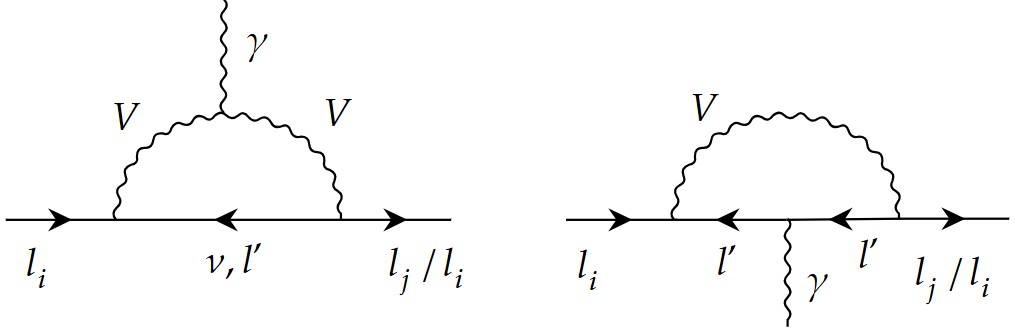}
	\caption{\label{muer}The Feynman diagrams of $l_i \to l_j \gamma$ and $g-2$.}
\end{figure*}

After discussing the constraints from tree-level processes, we now turn to the loop-induced processes, such as $l_i \to l_j \gamma$ and $g-2$ of charged leptons. We also derive constraints from lepton EDMs, which are predicted to be extremely small in the SM, while our model could potentially generate a significantly larger EDM to be detected. These processes are generated by similar diagrams, as shown in Figure \ref{muer} at one-loop level, due to the new interactions described in Eq. (\ref{f-v}), which can be rewritten as
\begin{eqnarray}
	\begin{aligned}
		\mathcal{L}=&-Y_{\alpha\beta}\left(\overline{l_{\beta}}\gamma^{\mu} P_R \nu_{\alpha}^{c}\right)V_{\mu}^{-}-Y_{\alpha\beta}^{*}\left(\overline{\nu_{\alpha}^{c}}\gamma^{\mu} P_R l_{\beta}\right)V_{\mu}^{+}\\&+\frac{Y_{\alpha\beta}}{2}\left(\overline{l_{\alpha}}\gamma^{\mu} P_L l_{\beta}^{c}-\overline{l_{\beta}}\gamma^{\mu} P_R l_{\alpha}^{c}\right)V_{\mu}^{--}+\frac{Y_{\alpha\beta}^{*}}{2}\left(\overline{l_{\beta}^{c}}\gamma^{\mu} P_L l_{\alpha}-\overline{l_{\alpha}^{c}}\gamma^{\mu} P_R l_{\beta}\right)V_{\mu}^{++}.
	\end{aligned}
	\label{LR}
\end{eqnarray}
When extending the SM by adding vector bosons, a key challenge is how to control the interactions to ensure the ultraviolet (UV) completeness so that loop corrections are finite and renormalizable. To address this issue, heavy vector particles typically emerge as additional gauge bosons associated with spontaneous symmetry breaking of a UV theory with a larger gauge group to the SM gauge group. We will take the condition $\kappa_w=\kappa_y=0$ in our study which may be obtained from larger gauge symmetry breaking~\cite{delAguila:2010mx, Biggio:2016wyy, Baker:2019sli} to ensure that no divergences arise in the calculations. Under this assumption, the coupling of $V$ with the photon is similar to the photon interaction with the W boson. 
Then the amplitude for process $l_i(p_1) \to l_j(p_2) \gamma(q)$ can be written as
\begin{eqnarray}
	iM=\overline{u}(p_2)(-ie\Gamma_{ji}^\mu )u(p_1)\tilde{\mathcal{A}}_{\mu}\left(-q\right),
\end{eqnarray}
where $-e$ is the electric charge of lepton and the vertex function can be written by
\begin{eqnarray}
	\label{llr}
	\Gamma_{ji}^\mu =i\sigma_{\mu\nu}q^\nu\frac{m_i}{2}\left[A_L^{l_il_j}P_L+A_R^{l_il_j}P_R\right],
\end{eqnarray}
where $P_{L,R}=\frac{1}{2}(1\mp\gamma_5)$. The decay width of $l_i \to l_j \gamma$ can be described by
\begin{eqnarray}
	\Gamma(l_i\to l_j\gamma)=\frac{\alpha_{\text{em}}}{4}m_{l_i}^5\left(|A_L^{l_il_j}|^2+|A_R^{l_il_j}|^2\right).
\end{eqnarray}
At the same time, we can also extract the lepton anomalous magnetic moment and EDM from the vertex as it is generated by similar diagrams in Figure \ref{muer}
\begin{eqnarray}
	\Delta a_l=m_l^2(A_L^{ll}+A_R^{ll}).
\end{eqnarray}
\begin{eqnarray}
	d_l=\frac{i e m_l}{2}(A_R^{ll}-A_L^{ll}).
\end{eqnarray}
From Eq. (\ref{LR}), we obtain~\cite{Lavoura:2003xp}
\begin{eqnarray}
	\begin{aligned}
		A_L^{l_il_j}=\frac{1}{16 \pi^2 m_i m_V^2}\left(\frac{5}{6}m_i(Y^\dagger Y)_{ij}+ \frac{7}{3}m_i(Y^\dagger Y)_{ij}+\frac{7}{3}m_j (Y Y^\dagger)_{ji}+6\sum \limits_{k}  m_k (Y_{kj}Y^*_{ik})
		\right),
	\end{aligned}
\end{eqnarray}
\begin{eqnarray}
	\begin{aligned}
		A_R^{l_il_j}=\frac{1}{16 \pi^2 m_i m_V^2}\left(\frac{5}{6}m_j(Y^\dagger Y)_{ij}+\frac{7}{3} m_i(Y Y^\dagger)_{ji}+\frac{7}{3}m_j (Y^\dagger Y)_{ij}+6 \sum \limits_{k}  m_k (Y_{jk}Y^*_{ki})
		\right),
	\end{aligned}
\end{eqnarray}
where $m_k$ is the mass of fermion in the middle of the loop diagram and the assumption that new vector mass is much bigger than the SM lepton mass is used.
So we can get the decay width of $l_i \to l_j \gamma$ with the influence of $V$
{\small
	\begin{eqnarray}
		\begin{aligned}
			\Gamma(l_i \to l_j \gamma)&=\frac\alpha4\frac{ m_i^5}{(16\pi^2)^2}\left[\left|\frac{\frac{19}{6}(Y^\dagger Y)_{ij}+6\sum \limits_{k} \frac{m_k}{m_i} Y_{k j}Y_{i k}^*}{ m_{V}^2}\right|^2+\left|\frac{\frac{7}{3}(Y Y^\dagger )_{ji}+6\sum \limits_{k} \frac{m_k}{m_i} Y_{j k}Y_{k i}^*}{ m_{V}^2}\right|^2\right],
		\end{aligned}
\end{eqnarray}}
where the condition that the initial-state particle mass is much bigger than the final-state particle mass is assumed. And the magnetic dipole moment and the electric dipole moment are
\begin{eqnarray}
	\Delta a_l=\frac{ m_l^2}{16\pi^2}\left[\left(\frac{\frac{19}{3}(Y^\dagger Y)_{l l}+\frac{14}{3}(Y Y^\dagger )_{l l}+12Re\sum \limits_{k} \frac{m_k}{m_{l}} Y_{k l}  Y_{l k}^*}{ m_{V}^2}\right)\right],
\end{eqnarray}
\begin{eqnarray}
	\label{edm}
	d_l=\sum \limits_{k}\frac{6 e  m_k }{16 \pi^2 }\frac{Im(Y_{kl} Y^*_{lk})}{m_V^2}.
\end{eqnarray}
Note that the initial and final state particles are identical here and only the combination of left-handed and right-handed couplings contributes to EDM, as can be seen from Eq. (\ref{LR}). The EDM of a lepton violates CP symmetry, which comes from the imaginary part of the coupling coefficients in our model, as shown in Eq. (\ref{edm}).

\begin{table}[htbp] 
	\footnotesize
	\begin{center}
		\begin{tabular}{lcl}
			\hline\hline
			Process & Experimental bound &  \multicolumn{1}{c}{Constraints}\\
			\hline 
			$\mu\to e\gamma$ & $4.2\times 10^{-13}$~\cite{MEG:2016leq} & $\begin{cases}
				\left|Y_{\mu e}Y_{\mu\mu}^{*}\right|<3.81\times10^{-11}\left(\frac{m_{V}}{\mathrm{GeV}}\right)^2, \left|Y_{e e}Y_{e \mu}^{*}\right|<1.10\times10^{-10}\left(\frac{m_{V}}{\mathrm{GeV}}\right)^2
				\\
				\left|Y_{\tau e}Y_{\mu\tau}^{*}\right|<3.46\times10^{-12}\left(\frac{m_{V}}{\mathrm{GeV}}\right)^2, \left|Y_{\tau e}Y_{\tau \mu}^{*}\right|<1.10\times10^{-10}\left(\frac{m_{V}}{\mathrm{GeV}}\right)^2
				\\
				\left|Y_{e e}Y_{\mu e}^{*}\right|<1.50\times10^{-10}\left(\frac{m_{V}}{\mathrm{GeV}}\right)^2, \left|Y_{e \mu}Y_{\mu \mu}^{*}\right|<4.19\times10^{-11}\left(\frac{m_{V}}{\mathrm{GeV}}\right)^2,
				\\
				\left|Y_{e \tau}Y_{\mu \tau }^{*}\right|<1.50\times10^{-10}\left(\frac{m_{V}}{\mathrm{GeV}}\right)^2, \left|Y_{e \tau}Y_{\tau \mu}^{*}\right|<3.46\times10^{-12}\left(\frac{m_{V}}{\mathrm{GeV}}\right)^2			
			\end{cases}$ \\
			$\tau\to e\gamma$ & $3.3\times 10^{-8}$~\cite{BaBar:2009hkt} & $
			\begin{cases}
				|Y_{ee} Y_{e \tau}^{*}|<7.34\times10^{-8}\left(\frac{m_{V}}{\mathrm{GeV}}\right)^2, |Y_{\mu e} Y_{\mu \tau}^{*}|<7.34\times10^{-8}\left(\frac{m_{V}}{\mathrm{GeV}}\right)^2
				\\
				|Y_{\tau e} Y_{\tau \tau}^{*}|<2.54\times10^{-8}\left(\frac{m_{V}}{\mathrm{GeV}}\right)^2, |Y_{\mu e} Y_{\tau \mu}^{*}|<6.52\times10^{-7}\left(\frac{m_{V}}{\mathrm{GeV}}\right)^2
				\\ 
				|Y_{e e} Y_{\tau e}^{*}|<9.85\times10^{-8}\left(\frac{m_{V}}{\mathrm{GeV}}\right)^2, |Y_{e \mu} Y_{\tau \mu}^{*}|<9.97\times10^{-8}\left(\frac{m_{V}}{\mathrm{GeV}}\right)^2
				\\
				|Y_{e \tau} Y_{\tau \tau}^{*}|<2.79\times10^{-8}\left(\frac{m_{V}}{\mathrm{GeV}}\right)^2, |Y_{e \mu} Y_{\mu \tau}^{*}|<6.52\times10^{-7}\left(\frac{m_{V}}{\mathrm{GeV}}\right)^2
			\end{cases}$ \\
			$\tau\to \mu\gamma$ & $4.2\times 10^{-8}$~\cite{Belle:2021ysv} & $\begin{cases}
				|Y_{e\mu} Y_{e \tau}^{*}|<8.28\times10^{-8}\left(\frac{m_{V}}{\mathrm{GeV}}\right)^2, |Y_{\mu \mu} Y_{\mu \tau}^{*}|<8.23\times10^{-8}\left(\frac{m_{V}}{\mathrm{GeV}}\right)^2
				\\
				|Y_{\tau \mu} Y_{\tau \tau}^{*}|<2.86\times10^{-8}\left(\frac{m_{V}}{\mathrm{GeV}}\right)^2, |Y_{\mu \mu} Y_{\tau \mu}^{*}|<1.11\times10^{-7}\left(\frac{m_{V}}{\mathrm{GeV}}\right)^2
				\\ 
				|Y_{e \mu} Y_{\tau e}^{*}|<1.52\times10^{-4}\left(\frac{m_{V}}{\mathrm{GeV}}\right)^2, |Y_{\mu e} Y_{\tau e}^{*}|<1.12\times10^{-7}\left(\frac{m_{V}}{\mathrm{GeV}}\right)^2
				\\
				|Y_{\mu \tau} Y_{\tau \tau}^{*}|<3.15\times10^{-8}\left(\frac{m_{V}}{\mathrm{GeV}}\right)^2, |Y_{\mu e} Y_{e \tau}^{*}|<1.52\times10^{-4}\left(\frac{m_{V}}{\mathrm{GeV}}\right)^2
			\end{cases} $\\
						$\Delta a_\mu(\mathrm{lattice})$ & $(107\pm69)\times10^{-11}$~\cite{Borsanyi:2020mff} & $\begin{cases}
				-1.58\times10^{-5}\left(\frac{m_{V}}{\mathrm{GeV}}\right)^2<Re(Y_{e \mu}Y_{\mu e}^{*})<5.37\times10^{-4}\left(\frac{m_{V}}{\mathrm{GeV}}\right)^2
				\\
				-4.56\times10^{-9}\left(\frac{m_{V}}{\mathrm{GeV}}\right)^2<Re(Y_{\tau \mu}Y_{\mu \tau}^{*})<1.55\times10^{-7}\left(\frac{m_{V}}{\mathrm{GeV}}\right)^2
				\\
				Y_{e \mu}Y_{e \mu }^{*}<4.92\times10^{-6}\left(\frac{m_{V}}{\mathrm{GeV}}\right)^2,				Y_{\tau \mu}Y_{\tau \mu }^{*}<4.92\times10^{-6}\left(\frac{m_{V}}{\mathrm{GeV}}\right)^2
				\\
				Y_{\mu e}Y_{\mu e }^{*}<6.68\times10^{-6}\left(\frac{m_{V}}{\mathrm{GeV}}\right)^2,				Y_{\mu \tau}Y_{\mu \tau }^{*}<6.68\times10^{-6}\left(\frac{m_{V}}{\mathrm{GeV}}\right)^2\\
				Y_{\mu \mu}Y_{\mu \mu }^{*}<1.36\times10^{-6}\left(\frac{m_{V}}{\mathrm{GeV}}\right)^2
			\end{cases} $ \\
				$\Delta a_e(\mathrm{Cs})$ &$	(-101\pm27)\times10^{-14}$ ~\cite{Fan:2022eto} & $\begin{cases}
				-3.54\times10^{-7}\left(\frac{m_{V}}{\mathrm{GeV}}\right)^2 <Re(Y_{\mu e}Y_{e \mu }^{*})<-1.38\times10^{-7}\left(\frac{m_{V}}{\mathrm{GeV}}\right)^2
					\\
				-2.11\times10^{-8}\left(\frac{m_{V}}{\mathrm{GeV}}\right)^2 <Re(Y_{\tau e}Y_{e \tau}^{*})<-8.20\times10^{-9}\left(\frac{m_{V}}{\mathrm{GeV}}\right)^2
				\end{cases} $ \\
				$\Delta a_e(\mathrm{Rb})$ &$	(34\pm16)\times10^{-14}$ ~\cite{Fan:2022eto} & $\begin{cases}
					1.87\times10^{-8}\left(\frac{m_{V}}{\mathrm{GeV}}\right)^2 <Re(Y_{\mu e}Y_{e \mu }^{*})<1.47\times10^{-7}\left(\frac{m_{V}}{\mathrm{GeV}}\right)^2
					\\
					1.11\times10^{-9}\left(\frac{m_{V}}{\mathrm{GeV}}\right)^2 <Re(Y_{\tau e}Y_{e \tau}^{*})<8.74\times10^{-9}\left(\frac{m_{V}}{\mathrm{GeV}}\right)^2
					\\
					7.33\times10^{-6}\left(\frac{m_{V}}{\mathrm{GeV}}\right)^2 <Y_{\mu e}Y_{\mu e }^{*}<5.76\times10^{-5}\left(\frac{m_{V}}{\mathrm{GeV}}\right)^2
					\\
					7.33\times10^{-6}\left(\frac{m_{V}}{\mathrm{GeV}}\right)^2 <Y_{	\tau e}Y_{\tau e }^{*}<5.76\times10^{-5}\left(\frac{m_{V}}{\mathrm{GeV}}\right)^2
					\\
					9.95\times10^{-6}\left(\frac{m_{V}}{\mathrm{GeV}}\right)^2 <Y_{e \mu}Y_{e \mu }^{*}<7.82\times10^{-5}\left(\frac{m_{V}}{\mathrm{GeV}}\right)^2
					\\
					9.95\times10^{-6}\left(\frac{m_{V}}{\mathrm{GeV}}\right)^2 <Y_{e \tau}Y_{e \tau }^{*}<7.82\times10^{-5}\left(\frac{m_{V}}{\mathrm{GeV}}\right)^2\\
					2.02\times10^{-6}\left(\frac{m_{V}}{\mathrm{GeV}}\right)^2 <Y_{e e}Y_{e e }^{*}<1.59\times10^{-5}\left(\frac{m_{V}}{\mathrm{GeV}}\right)^2
				\end{cases} $ \\
					$|d_e|$ &$	4.1\times10^{-30}\mathrm{~e~cm}$~\cite{Roussy:2022cmp}  & $\begin{cases}
					|Im(Y_{\mu e} Y^*_{e \mu})|<5.17\times10^{-14}\left(\frac{m_{V}}{\mathrm{GeV}}\right)^2, |Im(Y_{\tau e} Y^*_{e \tau})|<3.08\times10^{-15}\left(\frac{m_{V}}{\mathrm{GeV}}\right)^2
				\end{cases} $ \\
				$|d_\mu|$ &$	1.58\times10^{-19}\mathrm{~e~cm}$~\cite{Muong-2:2008ebm}  & $\begin{cases}
					|Im(Y_{e \mu} Y^*_{\mu e})|<4.12\times10^{-1}\left(\frac{m_{V}}{\mathrm{GeV}}\right)^2, |Im(Y_{\tau \mu} Y^*_{\mu \tau})|<1.19\times10^{-4}\left(\frac{m_{V}}{\mathrm{GeV}}\right)^2
				\end{cases} $ \\
				$|Re(d_\tau)|$ &$
					4.16\times10^{-18}\mathrm{~e~cm}	$ ~\cite{Belle:2021ybo} & $\begin{cases}
					|Im(Y_{e \tau} Y^*_{\tau e})|<1.08\times10^{1}\left(\frac{m_{V}}{\mathrm{GeV}}\right)^2,
					|Im(Y_{\mu \tau} Y^*_{\tau \mu})|<5.25\times10^{-2}\left(\frac{m_{V}}{\mathrm{GeV}}\right)^2
				\end{cases} $ \\
			\hline\hline
		\end{tabular}
		\caption{Experimental limit for Vector doublet. The second column shows the existing experimental limits. The inequalities are the restrictions that only considers the contribution of the corresponding parameters. The corresponding bounds are shown in the third column of the table at 90$\%$ CL.}
		\label{cons}
	\end{center}
\end{table}
From the experiment, we can also derive the limits to the coupling and mass of our new model. The most stringent experimental constraint on $l_i \to l_j \gamma$ is from $\mu \to e\gamma$~\cite{MEG:2016leq}, $
	{Br}(\mu\to e\gamma)<4.2\times10^{-13}~(90\% \mathrm{CL})$.
On the other hand, the muon $g-2$ measurements also exhibit a discrepancy with the SM prediction. The most recent measurement of the muon $g-2$ was conducted by the experiment at Fermilab~\cite{Muong-2:2023cdq}, which reported a new result $a_{\mu}(\exp)=116592059(22)\times10^{-11}$. Comparing this with the SM prediction~\cite{Aoyama:2020ynm}, we can obtain $\Delta a_\mu=(249\pm48)\times10^{-11}$, which deviates at the level of 5.1$\sigma$ from the SM prediction. 
However, the uncertainty in the leading-order hadronic contributions to muon $g-2$ dominates the theoretical uncertainty, and the results of recent lattice calculations are more supportive of this experimental
 value~\cite{Borsanyi:2020mff},	$\Delta a_\mu=(107\pm69)\times10^{-11}$.
We also have electron $g-2$ measurement~\cite{Fan:2022eto} but the discrepancy with the SM depends on the choice of experimental measurement of the fine-structure constant. If one uses the measurement from rubidium $\alpha(\mathrm{Rb})$~\cite{Morel:2020dww}, there is a $+2.2\sigma$ deviation $\Delta a_e(\mathrm{Rb})=(34\pm16)\times10^{-14}$.
However, if one uses the measurement from cesium $\alpha(\mathrm{Cs})$~\cite{Parker:2018vye}, 
there is a $-3.7 \sigma$ deviation	$\Delta a_e(\mathrm{Cs})=(-101\pm27)\times10^{-14}$. Even though we will show the limits from both $\Delta a_e$, one can see that the situation is less clear, meaning that the limits we obtained from electron $g-2$ require further confirmation through future experiments. One may also consider tauon $g-2$ to study the constraints. However, large uncertainties in the measurement of tauon $g-2$ have been reported ~\cite{ATLAS:2022ryk}, primarily due to the short lifetime of tauon. Therefore, the limitations of this experiment are not considered in this paper. What's more, numerous experimental programs are currently underway worldwide to probe the existence of non-zero EDMs in various systems. However, no evidence for EDMs has been observed to date. The most stringent constraint on lepton EDM is provided by the electron, with an upper bound of $|d_e|<4.1\times10^{-30}\mathrm{~e~cm}$~\cite{Roussy:2022cmp}. 

In Table \ref{cons} we show the limits obtained, assuming that all the coupling coefficients are of the same order of magnitude, and that all current experimental results are strictly satisfied at 90$\%$ CL. Note that the constraints in Table \ref{cons} correspond to the case where only one combination of parameters is considered to be non-zero at a time. One can see that the process $\mu \to e \gamma$ provides the most stringent constraints among $l_i \to l_j \gamma$ process. However, since the constrained parameters are not identical, all the restrictions are meaningful. It is also worth noting that we cannot determine the sign of constraints imposed by $\Delta a_\mu(\mathrm{lattice})$ at 90$\%$ CL and we cannot use $Y_{\mu e}Y_{\mu e }^{*}, Y_{	\tau e}Y_{\tau e }^{*}, Y_{e \mu}Y_{e \mu }^{*}, Y_{e \tau}Y_{e \tau }^{*}$ or $Y_{e e}Y_{e e }^{*}$ to explain $\Delta a_e(\mathrm{Cs})$ now as they cannot provide negative contribution. Moreover, the limits obtained from the electron $g-2$ do not fully overlap with those from other experiments, which calls for more precise measurements of the relevant parameters in future studies. For $Y_{e\mu}Y_{e\mu}^*$ and $Y_{\mu e}Y_{\mu e}^*$, the constraints given by $e^-e^+ \to l^-l^+$ scattering experiments are more stringent than the limits from $\Delta a_\mu(\mathrm{lattice})$ . We also note that $\Delta a_\mu(\mathrm{lattice})$ give a strong limit on $Y_{\mu\mu}Y_{\mu\mu}^*$ than neutrino trident scattering process. Additionally, the electron EDM provides the most stringent constraint on the imaginary part of coupling coefficients, while other limits are much weaker. However, improved experimental precision is expected to offer deeper insights into our model in the future.

\section{Conclusion} 

In this work, we investigate the theoretical framework and phenomenological implications of a new vector particle, $V$, which transforms as a doublet under the Standard Model gauge group with quantum numbers (1, 2, -3/2) and mediates $\Delta L =2$ processes. Our analysis focuses on the novel dimension-4 interactions of $V$ with SM bi-leptons and their potential signatures in various experimental processes. Using the latest experimental data, we derive stringent upper bounds on the couplings of $V$ to leptons.
	
The results suggest that process $\mu^- \to e^+e^-e^-$ provides the most stringent constraints in $l_\alpha^{-} \rightarrow l_\beta^{+} l_\rho^{-} l_\sigma^{-}$, which are $\left|Y_{e e} Y^{*}_{\mu e}\right| 
< 3.29 \times 10^{-11}\left(m_{V}/ \mathrm{GeV}\right)^{2}$, and $\left|Y_{e e} Y^{*}_{e \mu}\right| 
< 3.29 \times 10^{-11}\left(m_{V}/ \mathrm{GeV}\right)^{2}$. These constraints are crucial, as they directly probe the new interactions at tree-level and help to define the viable parameter space for the model.

Muonium and antimuonium oscillation is another important process we investigate. By comparing the predicted oscillation probability with the experimental upper bound, we derive a constraint $	{|Y_{\mu \mu}Y_{e e}^{*}|}<3.52\times10^{-8}\left(m_{V}/ \mathrm{GeV}\right)^{2}$. However, this constraint appears to be relatively weak.

Considering neutrino-lepton interactions, we can derive the limit $	{\left|Y_{\mu\mu}\right|^{2}}<2.46\times10^{-5}(m_{V}/ \mathrm{GeV})^2$ from the neutrino trident scattering process. Additionally, we obtain the limits $ |Y_{\mu\mu}Y^*_{ee}|<9.90\times10^{-6}~(\mathrm{CHARM})~(m_{V}/ \mathrm{GeV})^2 $ or $ |Y_{\mu\mu}Y^*_{ee}|<4.30\times10^{-6} ~(\mathrm{NuTeV})~(m_{V}/ \mathrm{GeV})^2$, where the limits come from the combination of the inverse muon decay $\nu_\mu e^-\to\mu^-\nu_e$ and the lepton number violating process $\bar{\nu}_\mu e^-\to\mu^-\bar{\nu}_e$.

Comparing with the experimental data from LEP II, we obtain constraints on the combination of different coupling coefficients. The most stringent limits are ${Y_{ee}Y_{ee}^*}<6.10\times10^{-6}\left(\frac{m_{V}}{ \mathrm{GeV}}\right)^{2}$, ${Y_{\mu e}Y_{\mu e}^*}<1.97\times10^{-7}\left(m_{V}/ \mathrm{GeV}\right)^2$, and ${Y_{\tau e}Y_{\tau e}^*}<2.68\times10^{-7}\left(m_{V}/ \mathrm{GeV}\right)^2$ in different $e^-e^+ \to l^-l^+$ processes.

Similarly, for the $l_i \to l_j \gamma$ process and anomalous $g-2$ of charged leptons, the experimental upper bounds lead to tight restrictions on the couplings and mass of the new vector doublet, as shown in Table \ref{cons}. The $\mu \to e \gamma$ process provides the best constraints $ \left|Y_{\tau e}Y_{\mu\tau}^{*}\right|<3.46\times10^{-12}\left(m_{V}/ \mathrm{GeV}\right)^2,$ and $ \left|Y_{e \tau}Y_{\tau \mu}^{*}\right|<3.46\times10^{-12}\left(m_{V}/ \mathrm{GeV}\right)^2$ among these one-loop processes. However, it is difficult to explain other experiment results and $g-2$ anomaly, especially the electron $g-2$, at the same time, as this would impose stringent constraints on the parameter space if the current experimental data are further confirmed. Additionally, experimental limits on EDMs impose significant constraints on the imaginary parts of coupling constants, with the electron EDM currently providing the most stringent bounds, $|Im(Y_{\mu e} Y^*_{e \mu})|<5.17\times10^{-14}\left(\frac{m_{V}}{\mathrm{GeV}}\right)^2$ and $ |Im(Y_{\tau e} Y^*_{e \tau})|<3.08\times10^{-15}\left(\frac{m_{V}}{\mathrm{GeV}}\right)^2$.

Some of the processes we studied provide much weaker constraints. It is worth mentioning that since the constrained parameter space are not identical, these bounds may still be useful in probing new physics model studied in this paper. Future high-precision experiments are expected to obtain more stringent constraints to provide deeper insights into the existence and properties of this novel vector particle.

\begin{acknowledgments}
The authors would like to thank Ming-Wei Li for helpful discussions and Jin Sun for valuable suggestions about the studies of inverse muon decay and EDM. This work is supported in part by the National Key Research and Development Program of China under Grant
No. 2020YFC2201501, by the Fundamental Research
Funds for the Central Universities, by National Natural Science Foundation of P.R. China (No.12090064,
12205063, 12375088 and W2441004).
\end{acknowledgments}

\end{document}